# Siberian forest fires: anomalies and trends from satellite data (2000-2019)

I. I. Mokhov [1,2], S. A. Sitnov [1]

[1]A.M. Obukhov Institute of Atmospheric Physics, Russian Academy of Sciences

[2]Lomonosov Moscow State University

mokhov@ifaran.ru

**Abstract**

The forest fires characteristics in Siberia detected by satellite data (MODIS instruments on Aqua and Terra platforms) for the period 2000-2019 are analyzed. Regional statistical data and distribution functions of wildfire characteristics are presented. Differences in spatio-temporal changes of forest fires of various intensity were revealed. An analysis of trends of the wildfire characteristics over the past two decades indicates an increase in the proportion of intense forest fires during the fire hazardous seasons, as well as an increase in the intensity of average Siberian wildfire in summer.

**Keywords:** forest fires, anomalies, trends, Siberia, satellite data

**Introduction**

Wildfires are among the most hazardous regional consequences of global warming. Biomass burning is an important source of various gaseous compounds and aerosol particles in the atmosphere, including toxic- and greenhouse gases (Andreae and Merlet 2001; Van der Werf et al. 2010; Shvidenko et al. 2011; Sitnov 2011; Eliseev et al. 2014; Sitnov and Mokhov 2017a, 2017b; Sitnov et al. 2017; Andreae 2019; Sitnov et al. 2020; Mokhov and Sitnov 2022). Aerosols associated with pyrogenic emissions change the radiative budget and thermal regime of the atmosphere (Mokhov and Gorchakova 2005; Gorchakova and Mokhov 2012; Gorchakov et al. 2014; Shukurov et al. 2014), thereby affecting the regional climate (Meehl et al. 2008). Once in the atmosphere, the biomass burning products worsen the quality of atmospheric air (Lei et al. 2013) and negatively affect human health (Janssen et al. 2012).

Global warming, which manifest itself most distinctly in the northern high latitudes, leads to changes in the long-term fire regimes in the boreal forests (Malevsky-Malevich et al. 2007; McKenzie et al. 2014). Wildfire observations in the North American forests indicates an increase in the annual number of wildfires and an earlier onset of the fire-hazardous period in recent decades (Westerling et al. 2006; Marlon et al. 2012; Abazoglou and Williams 2016). Under ongoing climate changes, in the boreal forests of Russia increase of the proportion of crown and underground fires are observed (Korovin 1996) with increase of the frequency of catastrophic wildfires (Shvidenko and Shchepaschenko 2013) and increase of the mean fire radiative power (Sitnov and Mokhov 2018). The formation of prolonged and intense fires is also associated with atmospheric blockings (Mokhov et al. 2020). According to model estimates, under global warming an increase in the frequency of atmospheric blockings, including the most prolonged ones, should be expected (Mokhov and Timazhev 2019; Mokhov 2020).

Siberian wildfires account for three quarters of forest fires in Northern Eurasia and more than half (about 55%) of the total number of forest fires throughout the boreal zone (Wooster and Zhang 2004; Sitnov and Mokhov 2018). A significant part of Siberian forests is in the Subarctic and Arctic zones. The entry of biomass burning products, including black carbon, directly into the high-latitude atmosphere makes severe Siberian forest fires an important factor of climate change in the Arctic.

The global and regional characteristics of wildfires, derived from satellite data was a subject of numerous studies (e.g. Dwyer et al. 2000; Kaufman et al. 2003; Giglio et al. 2006; Bondur et al. 2020; Mokhov et al. 2020). The aim of this work is to analyze the regional characteristics of Siberian forest fires and their tendencies over past two decades using satellite data. According to model estimates one should expect increasing the frequency of wildfires and the duration of the fire-hazardous period under global warming in the 21$^{st}$ century, in particular in the Asian part of Russia (Mokhov et al. 2006; Mokhov and Chernokulsky 2010; Mokhov 2022).

**Data analyzed**

An analysis of Siberian wildfires was carried out using active fire products obtained from the MODIS (MODerate resolution Imaging Spectroradiomer) instruments (Justice et al. 2002) installed onboard the Aqua and Terra satellites, residing at near-polar solar-synchronous orbits. MODIS measures radiances in the wavelength range from 0.4 to 14.6 μm in 36 spectral bands at different spatial resolutions (250, 500 and 1000 m, in nadir). The optical system of the instruments scans across the satellite's path the band of 2330 km wide, providing almost daily global coverage of the earth's surface by data at mid and high latitudes.

The number of hotspots (fire counts) and fire radiative power (FRP) detected by MODIS/Aqua during the period from July 2002 to December 2019 and those detected by MODIS/Terra from November 2000 to December 2019 were analyzed. MODIS detects fires in cloud-free conditions in ground pixels 1 km × 1 km in size (at nadir) using the context algorithm (Giglio et al. 2003). To identify thermal anomalies associated with forest fires (fire pixels), the threshold levels for radiation emitted at wavelengths of 4 μm and 11 μm are set. False fires are filtered using a comparative analysis of the brightness temperatures of the fire pixel and those of surrounding pixels and also by a comparative analysis of changes in the brightness temperatures of the fire pixel in the middle- and far IR ranges (Giglio et al. 2016). In this study the pixels characterized by a high confidence of fire detection (at least 80%) were only used.

The MODIS Active Fire Products data (C6, L2) processed by the standard algorithm (Giglio et al. 2016) were obtained through the FIRMS server (Fire Information for Resource Management System, https://earthdata.nasa.gov).

**Results and discussion**

The forest fires in the territory restricted by the coordinates 45-75°N and 60-140°E (Fig. 1, hereinafter referred to as Siberia) whose area is of about $15 \cdot 10^6$ km$^2$ were analyzed. Of about 20% of the considered territory is located north of the Arctic Circle, moreover the most part of Western Siberia (WS: 45-75°N, 60-90°E) and the whole of Eastern Siberia (ES: 45-75°N, 90-140°E) are located in the permafrost zone.

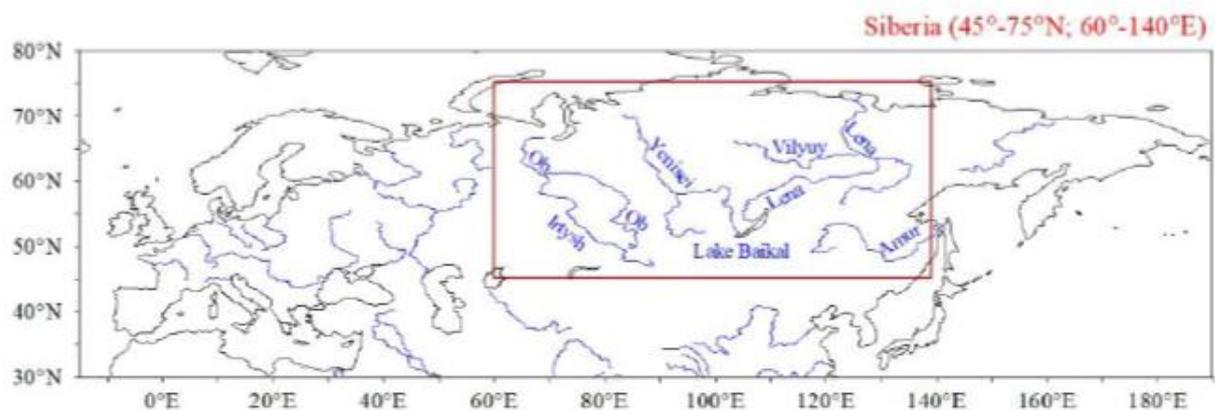

**Figure 1.** Northern Eurasia with Siberia characterized by a red rectangle.

On the territory of Siberia, tundra, forest-tundra, taiga and forest-steppe zones are successively distinguished from north to south. From west to east, Siberia is also geographically inhomogeneous. WS occupies the West Siberian Plain with the vast wetland area in its central part, which is occupied of about $8 \cdot 10^5$ km$^2$. In the forests of WS the dark coniferous tree species (spruce, fir and pine) predominate. In its turn, the most part of ES is occupied by the Central Siberian Plateau with a maximum relief height of up to 1701 m (Putorana Plateau), where a pronounced high-altitudinal zonality of vegetation takes place. Light coniferous tree species dominate in the forests of ES, mainly larch (The National Atlas of Russia, https://национальныйатлас.рф/cd2/english.html).

*Statistical characteristics of Siberian forest fires in 2000-2019*

Table 1 presents statistical characteristics of active fires detected by two MODIS instruments in the Siberian territory in the fire-hazardous seasons 2002-2019. The identical periods were chosen for a more correct comparison of the characteristics of fires detected by two MODIS instruments.

**Table 1.** Statistical characteristics of active fires, detected by the MODIS (Aqua/Terra) instruments in the territory of Siberia during the periods from April till September of 2002-2019.

| Statistics | MODIS/Aqua | MODIS/Terra |
|---|---|---|
| Fire counts | 724297 | 832891 |
| Mean FRP (MW) | 117 | 100 |
| Median FRP (MW) | 61 | 51 |
| Mode of FRP (MW) | 31 | 20 |
| SD of FRP (MW) | 193 | 176 |
| Kurtosis | 117 | 170 |
| Skewness | 8 | 9 |
| Maximum FRP (MW) | 8099 | 7899 |

The data presented in Table 1 indicate that the number of fires detected by MODIS/Terra exceeds those detected by MODIS/Aqua. Since MODIS diagnoses fires in cloud-free conditions, this circumstance often attributed to the increase (on average) in cloud cover in the afternoon, when the Aqua measurements occur (Giglio et al. 2006). However, the data analysis shows that this fact can be also due to the greater number of fire counts in the MODIS/Terra nighttime data. It can also be seen from Table 1 that the average, median and modal values of FRP of the Siberian wildfires obtained by MODIS/Aqua, are on the contrary larger than the same characteristics obtained by MODIS/Terra. In particular, the long-term average (median) FRP values according to the MODIS/Aqua instrument is 17% (20%) higher than the corresponding FRP characteristics obtained using MODIS/Terra data. The above differences can be attributed to the diurnal variation in the intensity of forest fires, whose maximum is reached in 1-2 hours in the afternoon (Andela et. 2015), because the daytime Aqua measurements occur near the daily maximum of wildfire intensity, while the Terra ones occur before noon, when the wildfires do not reach their maximum yet. The FRPs' characteristics from both MODIS instruments are characterized by approximately the same skewness, but substantially different kurtosis coefficients. According to MODIS/Aqua (Terra) data the most severe wildfire with FRP up to 8.1 (7.9) GW, was detected on 15 May 2008 (24 August 2015) in the ground pixel centered at 50.9°N, 108.0°E (53.3°N, 109.6°E), respectively.

Table 2 presents descriptive statistics of active fires detected during the fire-hazardous seasons for the entire period of MODIS/Terra observations (2000-2019). The statistics are presented for Siberia as a whole and separately for the territories of WS and ES. For comparison, the respective characteristics of wildfires for the territory of Eastern Europe (EE: 45-75°N, 20-60°E) are also shown (the last three regions were chosen to have the same area - of about 7.4·10$^6$ km$^2$).

**Table 2**. Statistical characteristics of wildfires detected by the MODIS/Terra instrument during the fire-hazardous seasons (from April till September) of 2000-2019 for Siberia as a whole (45-75°N, 60-140°E) and separately for Western Siberia (WS: 45-75°N, 60-100°E), Eastern Siberia (ES: 45-75°N, 100-140°E) and Eastern Europe (EE: 45-75°N, 20-60°E).

| Statistics | Siberia | WS | ES | EE |
|---|---|---|---|---|
| Fire counts | 868221 | 328949 | 539272 | 252544 |
| Mean FRP (MW) | 99 | 87 | 107 | 78 |
| Median FRP (MW) | 51 | 47 | 54 | 43 |
| Mode of FRP (MW) | 21 | 20 | 24 | 20 |
| SD of FRP (MW) | 174 | 140 | 192 | 117 |
| Kurtosis | 169 | 194 | 151 | 161 |
| Skewness | 9 | 10 | 9 | 9 |
| Maximum FRP (MW) | 7899 | 6749 | 7899 | 4940 |

Table 2 indicates that the number of active fires in ES is 1.6 (2.1) times greater than that in WS (EE), respectively. The above results can be explained by the fact that ES is the most densely forested territory of Northern Eurasia. It can be seen from Table 2 that the average (median) FRP values in ES is 23% (15%) higher than the corresponding values FRP in WS and that the above metrics in WS and ES exceed those in EE. The data in Table 2 also evidence that the modal FRP value for ES exceeds that for WS. Partially the above distinctions can be explained by the fact that in the WS forests larch prevails (https://национальныйатлас.рф/cd2/english.html), which is characterized by a higher calorific value than spruce, pine or fir, which dominate the ES forests. Characteristic differences between the average, median and modal values of FRP, as well as the high values of the skewness and kurtosis coefficients (Table 1,2) indicate that the frequency distributions of FRP are significantly different from normal one.

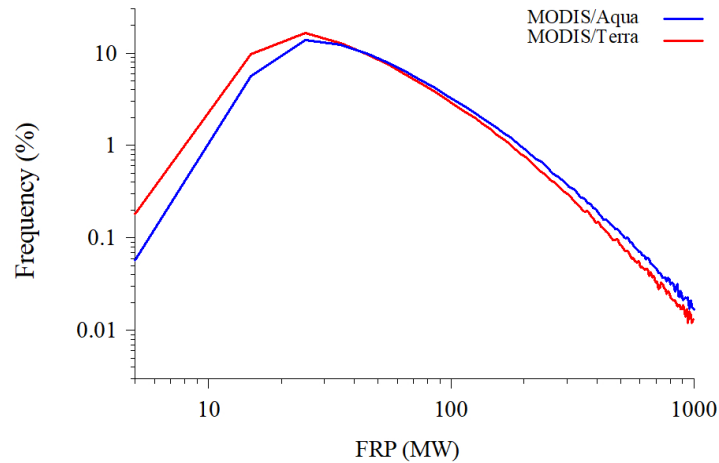

**Figure 2.** The distribution functions of the intensity of Siberian forest fires calculated using MODIS (Aqua/Terra) data. The presented range of FRP (0-1000 MW) includes more than 99% of total fire counts for each of the MODIS instruments.

The distribution functions (DFs) of FRP, obtained using MODIS/Aqua and MODIS/Terra data for the territory of Siberia (Fig. 2), reveal prevalence the weak fires in the total number of wildfires that leads to the positive skewness of DFs. Both DFs are unimodal, with the most probable values lying between 20 MW and 30 MW and reveal the sharp peaks, that explained the high positive values of kurtosis. A comparison of the DFs obtained using the data from two MODIS instruments indicates that the intense fires are more likely in the MODIS/Aqua data, whereas the weak ones are on the contrary more likely in the MODIS/Terra data. This discrepancy can be explained by the above-mentioned circumstance, namely the bulk of the wildfires detected by MODIS / Terra occur before the daily maximum of wildfires, while the fires detected by MODIS / Aqua occure near the daily wilfire maximum.

Spatial distributions of fires characterizing by different FRP values in the territory of Northern Eurasia during the fire-hazardous periods of 2001-2019 are presented in Fig. 3. The separation of data according to FRP gradations used in this study is rather arbitrary and principally intended to reveal the rough dependence of spatio-temporal characteristics of wildfires on their intensity. The dominance of relatively weak fires in the total mass of wildfires is confirmed by Fig. 3. Analysis of the data presented in Fig. 3 indicates that fires with an FRP less than 100 MW (Fig. 3a) account for three quarters of the total number of fires and more than a third of the total radiation power of fires in Northern Eurasia. Fires with FRP between 100 MW and 500 MW (Fig. 3b) account for almost a quarter of the total fire counts, making the significant contribution (about a half) to the total radiation power of wildfires. The wildfires characterized by FRP exceeding 500 MW (Fig 3c), being responsible for only 2% of the total fire counts, are contributing more than 20% to the total radiation power of forest fires in Northern Eurasia. It can be seen from Fig. 3c that the severe forest fires prevail in Eastern Siberia, with clusters of most intensive ones observed in Transbaikalia, as well as in the river valleys of Podkamennaya Tunguska, Angara, Vilyuy and Zeya (cf. Fig. 3c with Fig. 1).

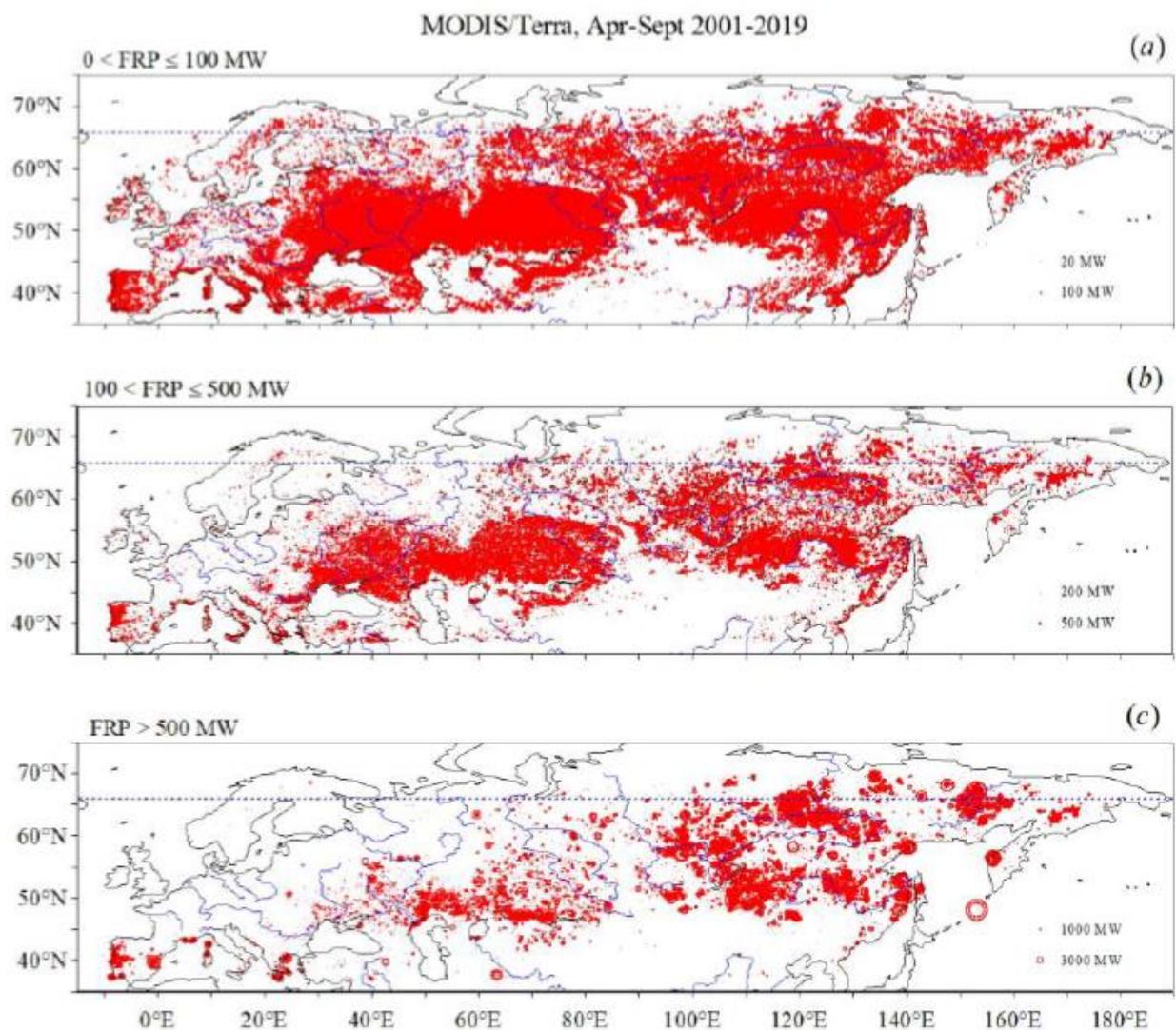

**Figure 3.** Spatial distributions of active fires of different intensity in the territory of Northern Eurasia detected by the MODIS/Terra instrument during the periods from April till September of 2001-2019 with FRP values less than 100 MW (a), between 100 and 500 MW (b), and exceeding 500 MW (c). The sizes of the red circles representing hotspots depend on FRP. The dotted blue lines denote Arctic Circle.

It should be noted that not all hotspots shown in Fig. 3 are forest fires. A small part of the hotspots are the gas torches in oil and gas production places (mainly in the territory of WS), while part is associated with volcanic activity in the Kamchatka Peninsula, Sakhalin Island and Kuril Islands. In particular, the intense hotspot centered at (48.1°N, 153.2°E) whose FRP exceeds 8.7 GW is associated with the eruption of Sarychev volcano (https://www.nasa.gov/multimedia/imagegallery/image_feature_1397.html) at Matua Island (North Kurils) on 12 June 2009.

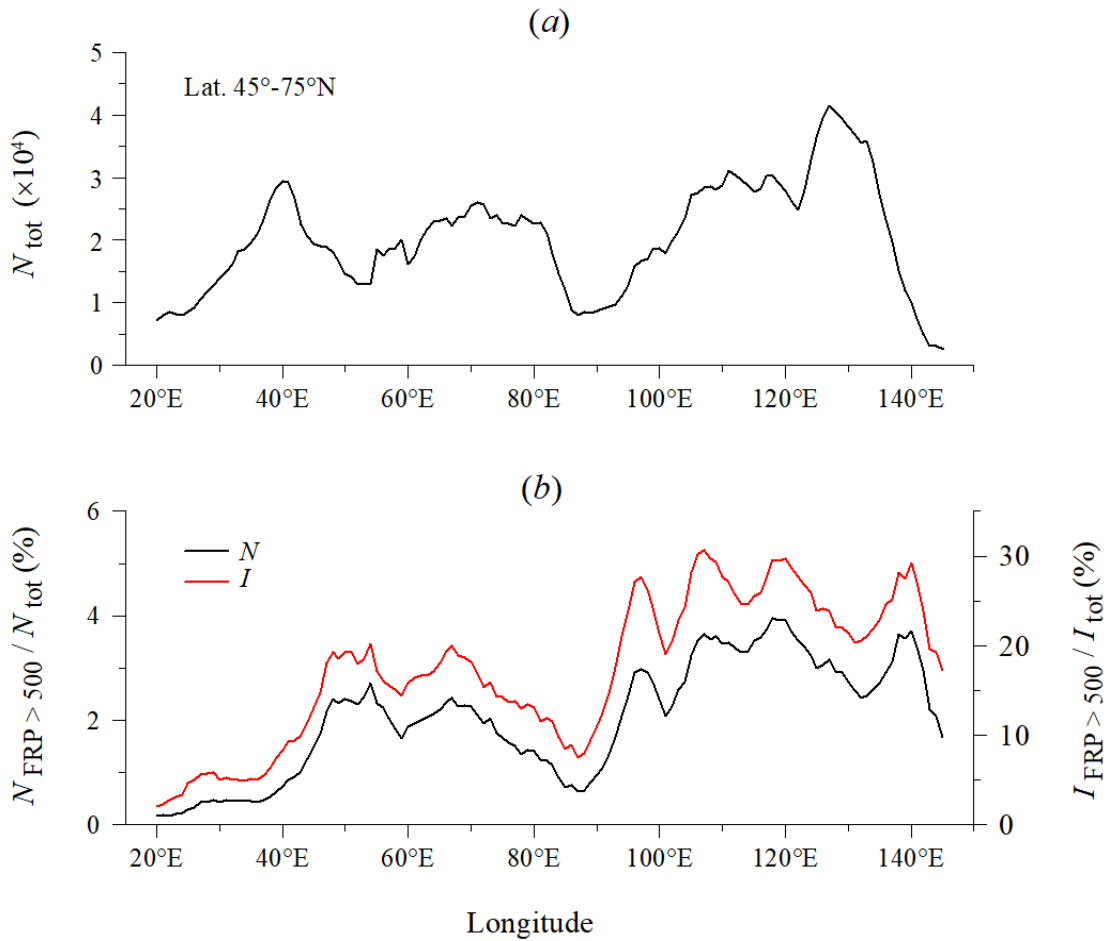

**Figure 4.** Longitudinal distribution of active fires detected within the latitudinal belt 45-75°N through the territory of Northern Eurasia in the fire-hazardous seasons of 2000-2019 (a); the same as (a) but for percentage ratios of the number (*N*) and cumulative FRP (*I*) of fires with FRP exceeding 500 MW to the total fire counts and total FRP, according to MODIS/Terra data (b).

The long-term longitudinal distribution of wildfires (Fig. 4a) principally reflects the orography and the density of forest cover in the territory of Northern Eurasia. The distribution characterized by maxima on the East European Plain (the longitudinal sector 35-45°E), on the West Siberian Plain (65-80°E) and on the Central Siberian Plateau (105-135°E). It can be seen from Fig. 4b, that from west to east of Northern Eurasia an increase in a number of intense wildfires is noticed. The ratio of the fire counts ($N$) with FRP exceeding 500 MW to the total fire counts ($N_{FRP>500} / N_{tot}$) for the territories EE, WS and ES in 2001-2019 is on average 0.5, 1.9, 2.6%, while the ratio of cumulative intensity ($I$) of fires with FRP exceeding 500 MW to the total fire intensity ($I_{FRP>500} / I_{tot}$) reaches 10, 16, 22%, respectively. The minimum in the longitudinal distribution of fires observed between 50°E and 60°E coincides with the Ural Mountains, stretching along the border of the East European and West Siberian plains from north to south for 2000 km. In the mountains, at altitudes above 800 m, sparse tundra vegetation dominates, which characterized by weak fire activity. Another minimum in the longitudinal distribution of wildfires located near 90°E coincides with the Yenisei River - the largest river system flowing to the Arctic Ocean. Yenisei is a natural border between Western and Eastern Siberia that revealed as a sharp change in the dominant type of vegetation. Spruce and fir forests which dominate the left bank of the Yenisei River are replaced by larch forests - on its right bank.

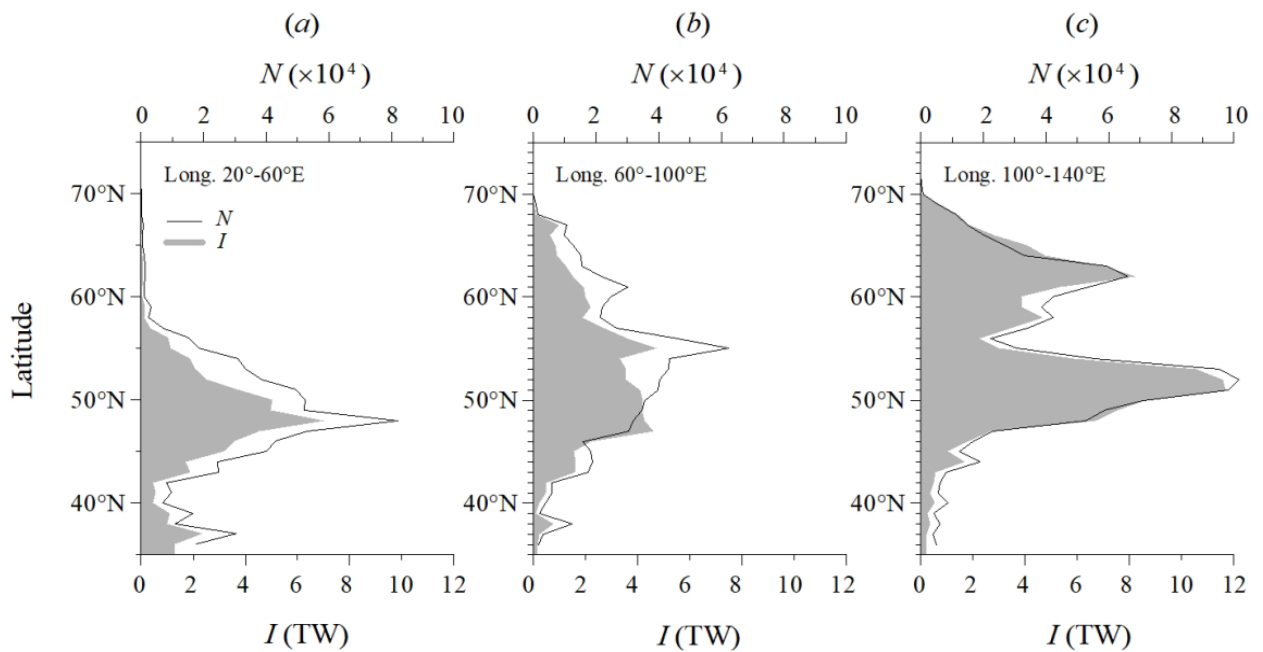

**Figure 5.** Latitudinal distributions of fire counts ($N$) and cumulative FRP ($I$) in different longitudinal sectors of Northern Eurasia: (a) Eastern Europe (20-60° E), (b) Western Siberia (60-100° E), (c) Eastern Siberia (100-140° E). MODIS/Terra data.

The long-term latitudinal distributions of wildfires in different subregions of Northern Eurasia are characterized by maxima at different latitudes (Fig. 5). These distributions also reflect the orography, type of woody vegetation and forest density cover in the region. In WS, the maximum number of wildfires is observed near the latitude 55°N (Fig. 5b), coinciding with a narrow latitudinal belt of subtaiga forests. In ES, the latitudinal distribution of wildfires reveals two maxima - one at latitudes 51-52°N and another one at 62°N with a deep minimum at 56°N (Fig. 5c). The maximum at 51-52°N coincides with the belt of forest-steppe and steppe vegetation, while that at 62°N falls on the latitudinal belt of larch forests on the Lena plateau. The minimum in the latitudinal distribution of wildfire characteristics coincides with the latitudinal location of the mountains characterized by tundra light forest. For the sake of comparison, Fig. 5a shows the latitudinal distribution of wildfires in EE. This distribution is characterized by an acute maximum on 48°N, indicating the prevalence of steppe fires in this region. Latitudinal distributions on Fig. 5 show that from west to east the wildfire region in Northern Eurasia tends to shift northward. The analysis results show that in the ES forests about 30% of fires occur at latitudes north of 60°N, that is in the subarctic and arctic zones. Analyzing the ratio of cumulative FRP ($I$) to fire counts ($N$), it can be inferred that the most intense fires in WS are observed in the latitudinal belt 46-50°N, while those in ES - in the latitudinal belts 48-50°N and 64-66°N.

*Seasonal cycle of Siberian wildfires*

Seasonal changes in the number of Siberian wildfires are shown in Fig. 6. The analysis of the presented results evidence that the 85% of the wildfires are observed from April to September (Fig. 6a). The seasonal dependences of the total number of Siberian wildfires (Fig. 6a) and the number of wildfires with FRP less than 100 MW (Fig. 6b) reveal two local maxima in July and May and local minimum in June. Unlike relatively weak fires the seasonal dependence of Siberian wildfires with FRP exceeding 500 MW characterized by only one summer maximum. Alongwith the seasonal cycle, the Box-Whisker plots also display the interannual variability of wildfires for individual months. It can be seen from Fig. 6, that regardless of FRP the greatest interannual variability of wildfires is observed in July and August.

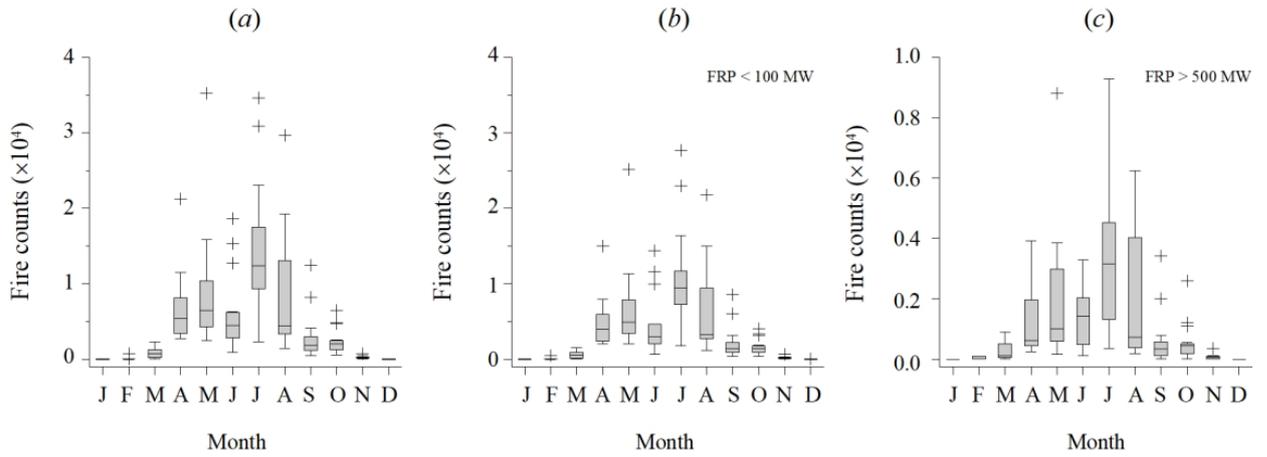

**Figure 6.** Box-Whiskers plots of the monthly fire counts (MODIS/Terra data) in the region (45-75° N, 60-140° E) during the period 2000-2019: (a) all data, (b) hotspots with FRP < 100 MW and (c) hotspots with FRP > 500 MW. Crosses denotes outliers.

The Box-Whiskers plots shown in Figs. 6 are based on median representation of data. Due to the strong positive asymmetry of the distribution of wildfire characteristics, the mean values are noticeably larger than the median ones. In particular, from April till September the mean value of fire counts is 1.4 times greater that corresponding median value. The maximum differences between means and medians are observed during the seasonal maximum of wildfires - from July to August. The mean value of total fire counts exceeds in August the median one by 2.2 times, moreover for the wildfires with FRP exceeding 500 MW this difference reaches 2.8.

*Trends of Siberian wildfires*

The time-series of the monthly regional mean fire counts ($N$) and the intensity ($I$) of Siberian wildfires indicates regular manifestation of fire activity in Siberian forests during the period of 2000-2019 with strong interannual and intra-annual variability of fire characteristics (Fig. 7a). It can be seen from Fig. 7a that the maxima of the fire activity in Siberian forests were noted in 2002, 2003, 2012, and 2019. Nominally, $N$ and $I$ are characterized by weak positive linear trends with the magnitudes 5 fires per month and 740 MW per month, respectively (Fig. 7a), which are not statistically significant, however.

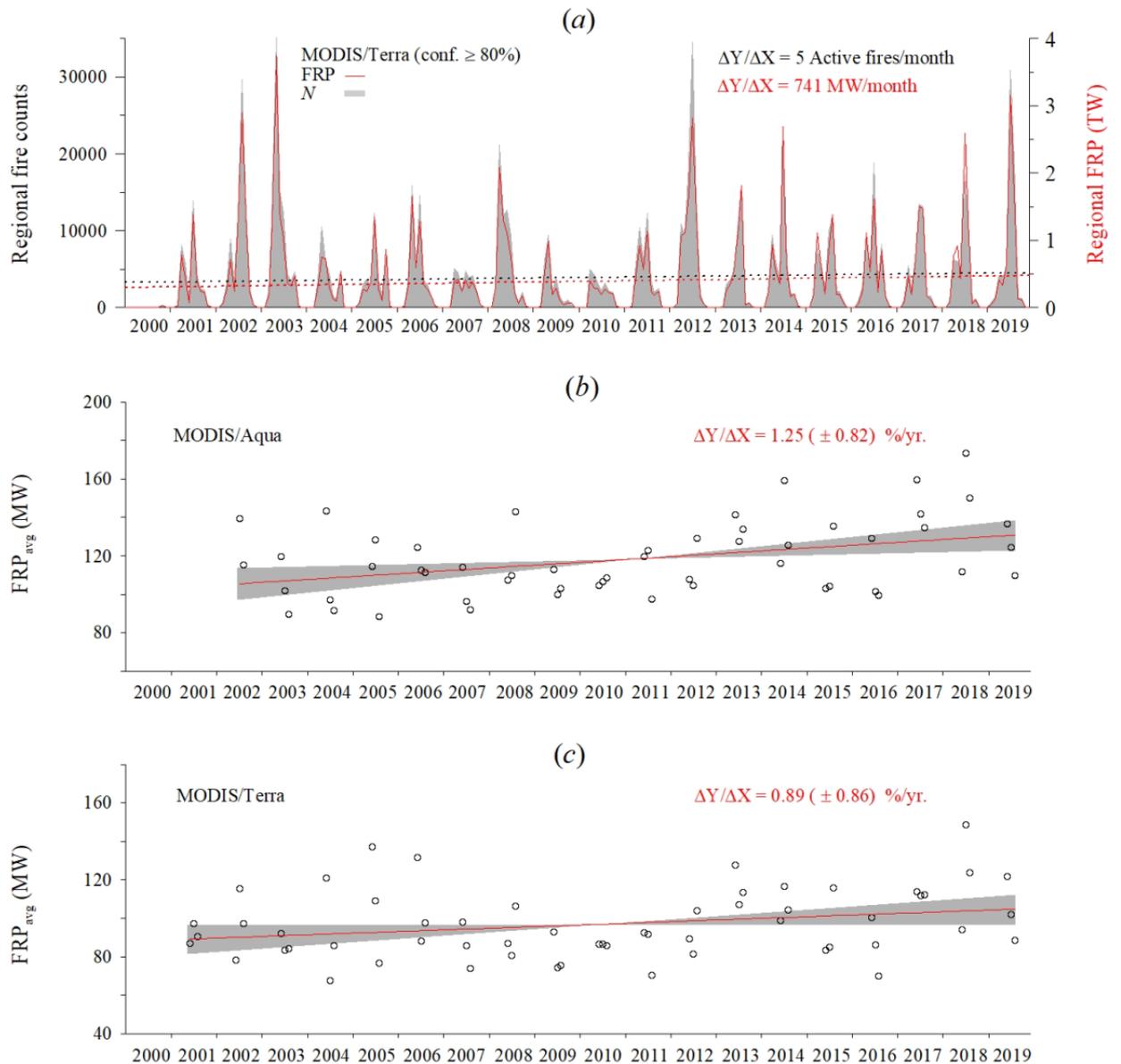

**Figure 7.** Monthly mean fire counts $N$ (grey area) and total FRP (red line) in the territory (45-75°N and 60-140°E) and their linear trends (dotted lines of corresponding colors) (a), average values of FRP for summer months and their linear trend (red line) with the 95% confidence intervals (shaded area), according to MODIS/Aqua data (b) and the same as (b), but according to MODIS/Terra data (c).

The results of the analysis revealed that the values of the coefficients of variation $N$ and $I$ (ratios of standard deviations to the corresponding averages, $C_v$) reach 150%. This fact evidences the heterogeneity of $N$ and $I$ sets, which makes impossible obtaining statistically valid characteristics of wildfires on the basis of used samples of $N$ and $I$, in particular, estimates of wildfires' trends. At the same time, the ratios of the total radiative power of wildfires in the region to the overall fire counts reveal significantly lower $C_v$ values, namely, $C_v = 16\%$ for MODIS/Terra data and $C_v = 19\%$ for MODIS/Aqua data that allows estimating the trends of above characteristic. According to the data from both MODIS instruments, the time series of the $I/N$ ratios reveal an increase in the period 2000-2019 (Fig. 7b,c). The above increase can be approximated by the statistically significant linear trends of $0.89 \pm 0.86$ % (MODIS/Terra) and $1.25 \pm 0.82$ % (MODIS/Aqua). It is useful to note that the ratio $I/N$ characterizes the radiation power of average Siberian wildfire ($FRP_{avg}$). Percentages were calculated in relation to corresponding mean values of $FRP_{avg}$ during the period 2000-2019 ($\pm 95\%$ confidence intervals of trends are also given). As follows from the above, both trends of $FRP_{avg}$ are statistically significant (at a $p$-level $\leq 0.05$). According to the MODIS/Aqua (Terra) instrument, in the summer periods of 2001 - 2019, the radiation power of an average Siberian wildfire grew by 25MW or 21% (16 MW or 17%), respectively. These facts can be associated with the growth of intense forest fires in Siberia.

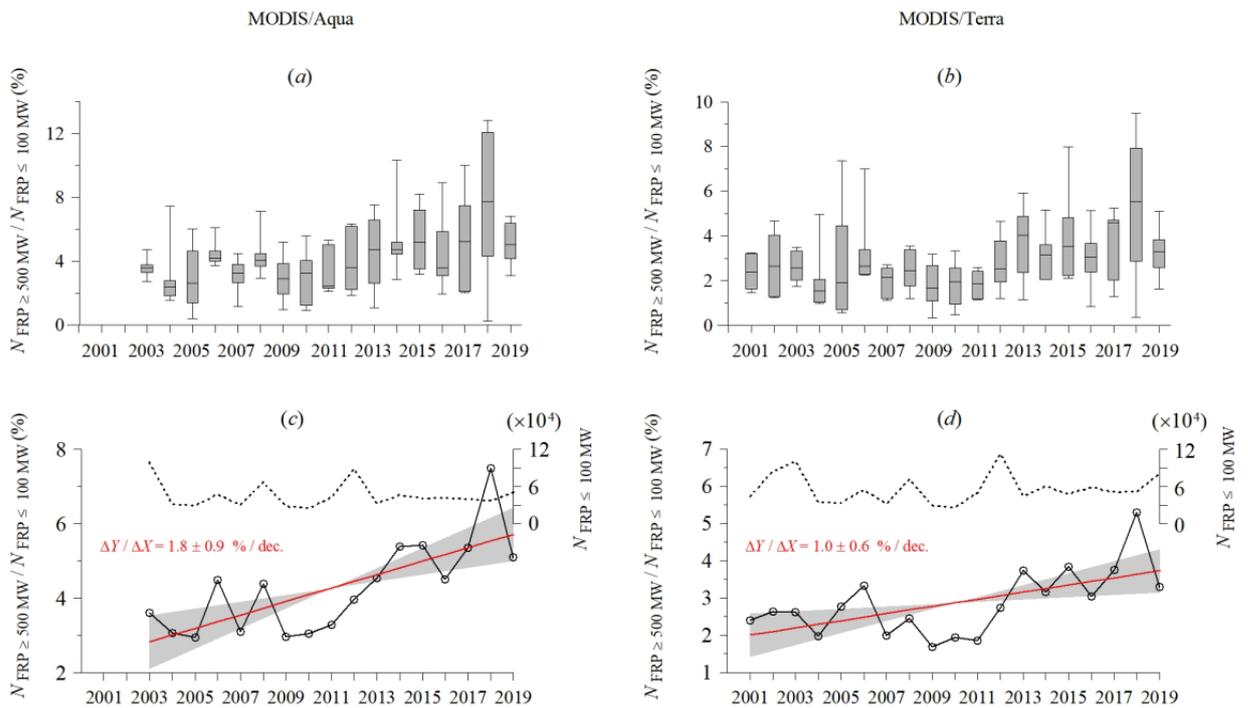

**Figure 8.** Box-Whiskers plot of ratio of the number of wildfires with FRP $\geq 500$ MW ($N_{FRP \geq 500}$) to those with FRP $\leq 100$ MW ($N_{FRP \leq 100}$) detected in the territory (45-75°N and 60-140°E) during the periods from April to September each year according to MODIS/Aqua data (a); (b) the same as (a), but according to MODIS/Terra data for; (c) the time series of the median values

of the ratio $N_{FRP \geq 500} / N_{FRP \leq 100}$ (black line) and its linear trend (red line) with the 95% confidence intervals (shaded area), according to MODIS/Aqua data; (d) the same as (c), but according to MODIS/Terra data. Also, in the plots (c) and (d) the fire counts with FRP ≤ 100 MW (dotted lines, right scale) are shown.

Figure 8a and Fig. 8b show interannual variations of the ratio ($R$) of the number of the Siberian wildfires with FRP exceeding 500 MW to those with FRP less than 100 MW during the fire-hazardous seasons of 2000-2019. The data analysis from both MODIS instruments, reveal an increase in $R$, accompanied by a certain tendency of increasing the dispersion of data with the maxima reached in 2018. The analysis results evidence that $R$ is characterized by positive statistically significant linear trend with angular coefficients of 1.8 ± 0.9 % / 10 years according to MODIS/Aqua (Fig. 8c) and 1.0 ± 0.6 % / 10 years according to MODIS / Terra (Fig. 8d). However, the increase in $R$ can be associated not only with the predominant increase of intense fires in the overall fire counts, but also with a decrease of a portion of weak fires during the considered period. An analysis of interannual variations of fires with a FRP less than 100 MW showed that long-term changes in the number of weak wildfires are characterized by weak and statistically insignificant linear trends (Figs. 8c and 8d). Thus, the revealed increase in $R$ is associated namely with the predominant increase in the number of intense Siberian forest fires during the period 2000-2019.

**Conclusions**

An analysis of characteristics of the Siberian forest fires obtained by the MODIS (Aqua, Terra) instruments during the period 2000-2019 revealed essential differences in the spatio-temporal dynamics of the wildfires of different intensity. The analysis results of longitudinal and latitudinal dependences of wildfires in the territory of Northern Eurasia indicate the predominance of the intense forest fires in Eastern Siberia, with significant part of them happen in the Subarctic and Arctic regions. The seasonal cycle of wildfires with FRP less than 100 MW, reveals spring and summer maxima, whereas that of wildfires with FRP exceeding 500 MW reveals only summer maximum. An analysis of the trends of characteristics of Siberian wildfires over the past 20 years indicates an increase of the radiation power of the average Siberian forest fire in the summer season as well as an increase a proportion of intense fires in the total number of Siberian wildfires.

**Acknowledgements**

We acknowledge the use of data products from the Land, Atmosphere Near real-time Capability for EOS (LANCE) system operated by NASA's Earth Science Data and Information System (ESDIS) with funding provided by NASA Headquarters. This work was supported by the Russian Science Foundation (project No. 19–17–00240).

**References**

Abatzoglou J.T., Williams A.P., 2016. Impact of anthropogenic climate change on wildfire across western US forests. Proc. NAS USA. 113, 11770-11775. https://doi.org/10.1073/pnas.1607171113

Andreae M.O., Merlet, P., 2001. Emission of trace gases and aerosols from biomass burning. Glob. Biogeochem. Cycl. 15, 955-966. doi: 10.1029/2000GB001382

Andreae M.O., 2019. Emission of trace gases and aerosols from biomass burning – an updated assessment. Atmos. Chem. Phys. 19, 8523–8546. https://doi.org/10.5194/acp-19-8523-2019

Andela N., Kaiser J.W., van der Werf G.R., Wooster M.J., 2015. New fire diurnal cycle characterizations to improve fire radiative energy assessments made from MODIS observations. Atmos. Chem. Phys. 15, 8831–8846. doi:10.5194/acp-15-8831-2015

Bondur V.G., Mokhov I.I., Voronova O.S., Sitnov S.A., 2020. Satellite monitoring of Siberian wildfires and their effects: Features of 2019 anomalies and trends of 20-year changes. Doklady Earth Sci. 492 (1), 370-375. doi: 10.1134/S1028334X20050049

Dwyer N, Pinnock S, Gregoire J.-M., Pereira J.M.C., 2000. Global spatial and temporal distribution of vegetation fires as determined from satellite observations. Intern. J. Remote Sens. 21, 1289–1302. doi: 10.1080/014311600210182

Eliseev A.V., Mokhov I.I., Chernokulsky A.V., 2014. Influence of ground and peat fires on $CO_2$ emissions into the atmosphere. Doklady Earth Sci. 459 (2), 1565-1569. doi: 10.1134/S1028334X14120034

Giglio L., Descloitres J., Justice C.O., Kaufman Y.J. 2003. An enhanced contextual fire detection algorithm for MODIS. Remote Sens. Environ. 87, 273-282. https://doi.org/10.1016/S0034-4257(03)00184-6

Giglio L., Csiszar I., Justice C.O. 2006., Global distribution and seasonality of active fires as observed with the Terra and Aqua Moderate Resolution Imaging Spectroradiometer (MODIS) sensors. J. Geophys. Res. 111, G02016. doi:10.1029/2005JG000142

Giglio L., Schroeder W., Justice C.O., 2016. The collection 6 MODIS active fire detection algorithm and fire products. Remote Sens. Environ. 178, 31-41. https://doi.org/10.1016/j.rse.2016.02.054

Gorchakov G.I., Sitnov S.A., Sviridenkov M.A., Semoutnikova E.G., Emilenko A. S., Isakov A.A., Kopeikin V.M., Karpov A.V., Gorchakova I.A., Verichev K.S., Kurbatov G.A., Ponomareva T.Ya., 2014. Satellite and ground-based monitoring of smoke in the atmosphere during the summer wildfires in European Russia in 2010 and Siberia in 2012. Intern. J. Remote Sens. 35, 5698-5721. doi: 10.1080/01431161.2014.945008

Gorchakova I.A., Mokhov I.I., 2012. The radiative and thermal effects of smoke aerosol over the region of Moscow during the summer fires of 2010. Izv., Atmos. Oceanic Phys. 48, 496-503. https://doi.org/10.1134/S0001433812050039

Janssen N.A.H., Gerlofs-Nijland M.E., Lanki T., Salonen R.O., Cassee F., Hoek G., Fischer P., Brunekreef B., Krzyzanowski M., 2012. Health effects of black carbon. WHO Rep. http://www.euro.who.int/__data/assets/pdf_file/0004/162535/e96541.pdf?ua=1

Justice C.O., Giglio L., Korontzi S., Owens J., Morisette J.T., Roy D., Descloitres J., Alleaume S., Petitcolin F., Kaufman Y., 2002. The MODIS fire products. Remote Sens. Environ. 83, 244-262. https://doi.org/10.1016/S0034-4257(02)00076-7

Kaufman Y., Ichoku C., Giglio L., Korontzi S., Chu D.A., Hao W.M., Li R.-R., Justice C.O., 2003. Fires and smoke observed from the Earth Observing System MODIS instrument—Products, validation, and operational use. Intern. J. Remote Sens. 24, 1765–1781. https://doi.org/10.1080/01431160210144741

Korovin G.N. 1996. Analysis of distribution of forest fires in Russia. Fires in Ecosystems of Boreal Eurasia, J.G. Goldammer and V.V. Furyaev (eds.), Springer, Dordrecht, 112-128. https://doi.org/10.1007/978-94-015-8737-2_8

Lei W., Li G., Molina L.T., 2013. Modeling the impacts of biomass burning on air quality in and around Mexico City. Atmos. Chem. Phys. 13, 2299–2319. https://doi.org/10.5194/acp-13-2299-2013

Malevskii-Malevich S.P., Mol'kentin E.K., Nadezhina E.D., Semioshina A.A., Sall' I.A., Khlebnikova E.I., Shklyarevich O.B., 2007. Analysis of changes in fire-hazard conditions in the forests in Russia in the 20th and 21st centuries on the basis of climate modeling. Rus. Meteorol. Hydrol. 32, 154–161. https://doi.org/10.3103/S1068373907030028

Marlon J.R,, Bartlein P.J., Gavin D.G., Long C.J., Anderson R.S., Briles C.E., Brown K.J., Colombaroli D., Hallett D.J., Power M.J., Scharf E.A., Walsh M.K., 2012. Long-term perspective on wildfires in the western USA. PNAS 109, E535-E543. https://doi.org/10.1073/pnas.1112839109

McKenzie D., Shankar U., Keane R.E., Stavros E.N., Heilman W.E., Fox D.G., Riebau A.C., 2014. Smoke consequences of new wildfire regimes driven by climate change. Earth's Fut. 2, 35–59. doi:10.1002/2013EF000180

Meehl G.A., Arblaster J.M., Collins W.D., 2008. Effects of black carbon aerosols on the Indian monsoon. J. Climate. 21, 2869 – 2882. https://doi.org/10.1175/2007JCLI1777.1

Mokhov I.I., 2020. Russian climate research in 2015–2018. Izvestiya, Atmos Oceanic Phys. 56 (4), 325-343. https://doi.org/10.1134/S0001433820040064

Mokhov I.I., 2022. Climate change: Causes, risks, consequences, and problems of adaptation and regulation. Herald Rus. Acad. Sci. 92 (1), 1–11.

https://doi.org/10.1134/S101933162201004X

Mokhov I.I., Bondur V.G., Sitnov S.A., Voronova O.S., 2020. Satellite monitoring of wildfires and emissions into the atmosphere of combustion products in Russia: Relation to atmospheric blockings. Doklady Earth Sci. 495, 921–924.

https://doi.org/10.1134/S1028334X20120089

Mokhov I.I., Chernokulsky A.V., 2010. Regional model assessments of forest fire risks in the Asian part of Russia under climate change. Geogr. Nat. Resources 31 (2), 165-169. https://doi.org/10.1016/j.gnr.2010.06.012

Mokhov I.I., Chernokulsky A.V., Shkolnik I.M., 2006. Regional model assessments of fire risks under global climate changes. Doklady Earth Sci. 411A (9), 1485-1488.

https://doi.org/10.1134/S1028334X06090340

Mokhov I.I., Gorchakova I.A., 2005. Radiation and temperature effects of summer fires in 2002 in the Moscow region. Doklady Earth Sci. 400, 160-163.

Mokhov I.I., Sitnov S.A., 2022. Siberian wildfires and related gas and aerosol emissions into the atmosphere over the past decades. https://arxiv.org/abs/2206.06999

Mokhov I.I., Timazhev A.V., 2019. Atmospheric blocking and its frequency in the 21st century simulated with the ensemble of climate models. Rus. Meteorol. Hydrol. 44 (6), 369-377.

Shvidenko A.Z., Schepaschenko D.G., 2013. Climate change and wildfires in Russia. Contemp. Probl. Ecol. 6, 683–692. https://doi.org/10.1134/S199542551307010X

Shvidenko A.Z., Shchepashchenko D.G., Vaganov E.A., Sukhinin A.I., Maksyutov S.S., McCallum I., Lakida I.P., 2011. Impact of Wildfire in Russia between 1998–2010 on Ecosystems and the Global Carbon Budget. Doklady Earth Sci. 441, 1678-1682. doi: 10.1134/S1028334X11120075

Shukurov K.A., Mokhov I.I., Shukurova L.M., 2014. Estimate for radiative forcing of smoke aerosol from 2010 summer fires based on measurements in the Moscow Region. Izv., Atmos. Oceanic Phys. 50 (3), 256-265.

Sitnov S.A., 2011. Aerosol optical thickness and the total carbon monoxide content over the European Russia territory in the 2010 summer period of mass fires: Interrelation between the variation in pollutants and meteorological parameters. Izv., Atmos. Oceanic Phys. 47, 714–728. doi: 10.1134/S0001433811060156

Sitnov S.A., Mokhov I.I., 2017a. Formaldehyde and nitrogen dioxide in the atmosphere during summer weather extremes and wildfires in European Russia in 2010 and Western Siberia in 2012. Intern. J. Remote Sens. 38, 4086-4106. doi: 10.1080/01431161.2017.1312618

Sitnov S.A., Mokhov I.I., 2017b. Anomalous transboundary transport of the products of biomass burning from North American wildfires to Northern Eurasia. Doklady Earth Sci. 475, 832–835. doi: 10.1134/S1028334X17070261

Sitnov S.A., Mokhov I.I., Gorchakov G.I., Dzhola A.V., 2017. Smoke haze over the European part of Russia in the summer of 2016: A link to wildfires in Siberia and atmospheric circulation anomalies. Rus. Meteorol. Hydrol. 42, 518-528. doi: 10.3103/S1068373917080052

Sitnov S.A., Mokhov I.I., 2018. Comparative analysis of active fire characteristics in the boreal forests of Eurasia and North America based on satellite data. Izv., Atmos. Oceanic Phys. 54, 966-978. doi: 10.1134/S0001433818090347

Sitnov S.A., Mokhov I.I., Likhosherstova A.A., 2020. Exploring large-scale black-carbon air pollution over Northern Eurasia in summer 2016 using MERRA-2 reanalysis data. Atmos. Res. 235, 104763. doi: 10.1016/j.atmosres.2019.104763

Van der Werf G.R., Randerson J.T., Giglio L., Collatz G.J., Mu M., Kasibhatla P.S., Morton D.C., DeFries R.S., Jin Y., van Leeuwen T.T., 2010. Global fire emissions and the contribution of deforestation, savanna, forest, agricultural and peat fires (1997–2009). Atmos. Chem. Phys. 10, 11707-11735. https://doi.org/10.5194/acp-10-11707-2010

Westerling A.L., Hidalgo H.G., Cayan D.R., Swetnam T.W., 2006. Warming and earlier spring increase western U.S. forest wildfire activity. Science 313, 940-943. doi:10.1126/science.1128834

Wooster M.J., Zhang Y.H., 2004. Boreal forest fires burn less intensely in Russia than in North America. Geophys. Res. Lett. 31, L20505. doi:10.1029/2004GL020805